%% file: DAC2024-main.tex
\documentclass[sigconf,natbib=false]{acmart}
\renewcommand\footnotetextcopyrightpermission[1]{} % removes footnote with conference information in first column
\AtBeginDocument{%
  }
\newcommand{\tabincell}[2]{\begin{tabular}{@{}#1@{}}#2\end{tabular}} 
\RequirePackage[
  datamodel=acmdatamodel,
  style=acmnumeric,
  sorting=none
  ]{biblatex}

\usepackage{makecell}
\usepackage{multirow}
\usepackage{booktabs}
\usepackage{footnote}
\usepackage{threeparttable}
\begin{document}

%%
%% The "title" command has an optional parameter,
%% allowing the author to define a "short title" to be used in page headers.
\title{DEFA: Efficient Deformable Attention Acceleration via Pruning-Assisted Grid-Sampling and Multi-Scale Parallel Processing}

%%
%% The "author" command and its associated commands are used to define
%% the authors and their affiliations.
%% Of note is the shared affiliation of the first two authors, and the
%% "authornote" and "authornotemark" commands
%% used to denote shared contribution to the research.

\author{Yansong Xu$^{1}$, Dongxu Lyu$^{1}$, Zhenyu Li$^{1}$, Zilong Wang$^{1}$, Yuzhou Chen$^{1}$, Gang Wang$^{1}$, Zhican Wang$^{1}$, Haomin Li$^{1}$, Guanghui He$^{1,2}$}
\affiliation{%
  \institution{
  $^{1}$School of Electronic Information and Electrical Engineering, Shanghai Jiao Tong University, Shanghai, China\\
  $^2$MoE Key Lab of Artificial Intelligence, AI Institute, Shanghai Jiao Tong University, China}
}
% \email{{xys-13, lvdongxu, ambitious-lzy, wangzilongsjtu, huygens, wangganginJSTU, Zhican_wang, haominli, guanghui.he}@sjtu.edu.cn}
\email{{xys-13, lvdongxu, ambitious-lzy, wangzilongsjtu, huygens, wangganginJSTU, guanghui.he}@sjtu.edu.cn}
% \email{Email: {xys-13, guanghui.he}@sjtu.edu.cn}
\authornote{Corresponding author: Guanghui He.}

\begin{abstract}
% Multi-scale deformable attention has emerged as a key mechanism in various vision tasks, demonstrating superior performance  than traditional attention-based solutions.
% Deformable attention has emerged as a key mechanism in various vision tasks, demonstrating superior performance with explicitly linear complexity.

% Multi-scale deformable attention (MSDeformAttn) has become a key mechanism in various vision tasks for multi-scale grid-sampling.
% However, this operator incurs irregular data access and enormous memory requirement, causing severe PE under-utilization.
% Meanwhile, existing approaches for attention acceleration cannot be directly applied to MSDeformAttn due to lack of support for this distinct procedure. 
% Therefore, we propose a dedicated algorithm-architecture co-design dubbed DEFA, the first-of-its-kind method for MSDeformAttn acceleration.
% At the algorithm level, DEFA adopts frequency-weighted pruning for feature maps and probability-aware pruning for sampling points, alleviating memory footprint by over 80\%.
% At the architecture level, it explores the multi-scale parallelism to boost the throughput significantly and further reduces memory access via fine-grained operator fusion and feature map reuse.
% DEFA achieves 10.1-31.9$\times$ speedup and 20.3-37.7$\times$ energy savings over GPUs. 
% It also rivals the related accelerators by 2.2-3.7$\times$ energy efficiency improvement.

Multi-scale deformable attention (MSDeformAttn) has emerged as a key mechanism in various vision tasks, demonstrating explicit superiority attributed to multi-scale grid-sampling.
However, this newly introduced operator incurs irregular data access and enormous memory requirement, leading to severe PE under-utilization.
Meanwhile, existing approaches for attention acceleration cannot be directly applied to MSDeformAttn due to lack of support for this distinct procedure. 
Therefore, we propose a dedicated algorithm-architecture co-design dubbed DEFA, the first-of-its-kind method for MSDeformAttn acceleration.
At the algorithm level, DEFA adopts frequency-weighted pruning and probability-aware pruning for feature maps and sampling points respectively, alleviating the memory footprint by over 80\%.
At the architecture level, it explores the multi-scale parallelism to boost the throughput significantly and further reduces the memory access via fine-grained layer fusion and feature map reusing.
Extensively evaluated on representative benchmarks, DEFA achieves  10.1-31.9$\times$ speedup and 20.3-37.7$\times$ energy efficiency boost compared to powerful GPUs. 
It also rivals the related accelerators by 2.2-3.7$\times$ energy efficiency improvement while providing pioneering support for MSDeformAttn.

\end{abstract}
\keywords{Transformer, Deformable Attention, Pruning, Domain-Specific Acceleration, Grid-Sampling}

% \received{20 February 2007}
% \received[revised]{12 March 2009}
% \received[accepted]{5 June 2009}

%%
%% This command processes the author and affiliation and title
%% information and builds the first part of the formatted document.
\maketitle

\input{1_intro}

\input{2_pre}
\input{3_algo}
\input{4_arch}

\input{5_eval}
\input{6_conclusion}

%%
%% The acknowledgments section is defined using the "acks" environment
%% (and NOT an unnumbered section). This ensures the proper
%% identification of the section in the article metadata, and the
%% consistent spelling of the heading.
\begin{acks}
This work was supported by the National Natural Science Foundation of China under Grant 62074097.
\end{acks}

%%
%% Print the bibliography
%%
\printbibliography

%%
%% If your work has an appendix, this is the place to put it.
% \appendix

% \section{Research Methods}

% \subsection{Part One}

% Lorem ipsum dolor sit amet, consectetur adipiscing elit. Morbi
% malesuada, quam in pulvinar varius, metus nunc fermentum urna, id
% sollicitudin purus odio sit amet enim. Aliquam ullamcorper eu ipsum
% vel mollis. Curabitur quis dictum nisl. Phasellus vel semper risus, et
% lacinia dolor. Integer ultricies commodo sem nec semper.

% \subsection{Part Two}

% Etiam commodo feugiat nisl pulvinar pellentesque. Etiam auctor sodales
% ligula, non varius nibh pulvinar semper. Suspendisse nec lectus non
% ipsum convallis congue hendrerit vitae sapien. Donec at laoreet
% eros. Vivamus non purus placerat, scelerisque diam eu, cursus
% ante. Etiam aliquam tortor auctor efficitur mattis.

% \section{Online Resources}

% Nam id fermentum dui. Suspendisse sagittis tortor a nulla mollis, in
% pulvinar ex pretium. Sed interdum orci quis metus euismod, et sagittis
% enim maximus. Vestibulum gravida massa ut felis suscipit
% congue. Quisque mattis elit a risus ultrices commodo venenatis eget
% dui. Etiam sagittis eleifend elementum.

% Nam interdum magna at lectus dignissim, ac dignissim lorem
% rhoncus. Maecenas eu arcu ac neque placerat aliquam. Nunc pulvinar
% massa et mattis lacinia.

\end{document}

%% file: 1_intro.tex
\section{Introduction}

% DEtection TRansformer (DETR) has gained an increasing popularity in object detection \cite{detr}. 
% It eliminates the need for hand-crafted components like non-maximum suppression (NMS) post-processing and builds a fully end-to-end detector based on an encoder-decoder transformer architecture, achieving a promising performance. 
% Despite its innovative architecture and superior performance, DETR is still struggling to converge and detect small objects. 
DEtection TRansformer (DETR) has gained increasing popularity in object detection due to the promising performance from the end-to-end optimizable network architecture.
Recently, multi-scale deformable attention (MSDeformAttn) \cite{zhu2021deformable}, inspired by deformable convolution (DeformConv) \cite{deformconv}, is proposed to further improve the DETR resolution on small objects with linear complexity, which only samples a small set of key points from multi-scale feature maps (fmaps) instead of traversing across all of them via $\mathcal{O}(n^2)$-level $\mathbf{Q} \times \mathbf{K}^{T}$ in traditional attention \cite{detr}.
Benefiting from multi-scale grid-sampling (MSGS), Deformable DETR achieves state-of-the-art detection accuracy so that MSDeformAttn has been widely adopted in recent 2D \cite{dndetr, dino,zhu2021deformable} and 3D \cite{bevformer,voxformer,tpvformer} object detection networks.
However, MSDeformAttn suffers from significant computation inefficiency on general-purpose platforms such as CPUs and GPUs.
% The inference framerate (fps) of Deformable DETR on Nvidia RTX 3090Ti GPU is less than 10 fps, which makes it hard to meet the real-time requirement. 
% For instance, Deformable DETR\cite{zhu2021deformable} executes on powerful Nvidia RTX 3090Ti GPU with only 9.7fps, far away from the real-time requirement in various tasks.
For instance, Deformable DETR (173GFLOPs) \cite{zhu2021deformable} executes at only 9.7fps even on a powerful Nvidia RTX 3090Ti GPU, while Faster R-CNN\cite{FasterR-CNN} with a similar workload (180GFLOPs) can reach over 25fps for the same task. 
With our deep analysis, MSDeformAttn takes up to 54.7\% of end-to-end inference latency while the MSGS procedure dominates (over 60\%) within each attention layer, which becomes the major efficiency bottleneck. 
% The results show that the grid-sample procedure leads to inefficiency.
Suffering from the dynamically unordered and unbounded sampling candidates all over the feature maps, grid-sampling results in heavily irregular memory access and high cache miss rate, further leading to severe PE under-utilization on GPUs. 
% Therefore, the general-purpose platforms cannot rely on the principle of locality, and thereby encounter low utilization of processing units and high missing rate of cache. 
In addition, applying multi-scale feature maps increases both the number of sampling points and the size of sampled fmaps by 21.3$\times$ compared to using single-scale ones, exacerbating the intensity of memory footprint. 
 % Unpredictable sampling on multi-scale feature maps demands an impracticable 19.6MB on-chip buffer. 
%(1. xx ms needed to be marked here to clarify it cannot meet the real-time requirement; 2. grid-sampling is non-uniform and results in low utilization of a processing unit; 3. large-size feature maps and random sampling cause heavy memory access consumption and huge on-chip memory requirement (like 200MB or 10GB)) xxxxxx xxxxxx xxxxxx xxxxxx xxxxxx xxxxxx xxxxxx xxxxxx xxxxxx xxxxxx xxxxxx xxxxxx xxxxxx.

Domain-specific accelerator (DSA) is an effective solution to improve the processing efficiency on resource-limited terminals. 
% The existing DSAs that optimize either self-attention \cite{attention} or deformable convolution \cite{deformconv} cannot maintain superior improvement on MSDeformAttn. 
Many existing DSAs \cite{spattn2021, elsa, qinghua2023} have been proposed to optimize attention \cite{detr} via reducing redundant computation between weak-related tokens. 
Nevertheless, MSDeformAttn reconstructs vastly different attention pipelines based on MSGS so that these works cannot maintain the superiority of MSDeformAttn due to a lack of support for grid-sample. 
Although some works \cite{huang2021codenet, li2022computational} also enhance similar grid-sample-based operators like deformable convolution\cite{deformconv}, their methods, especially the aggressive sampling range restriction induce unacceptable accuracy loss and also cannot be applied to MSDeformAttn dataflow directly.
% Although some works \cite{huang2019algorithm, huang2021codenet, ahn2020efficient, li2022computational} also enhance similar grid-sample-based operators like deformable convolution\cite{deformconv}, their methods, especially the aggressive sampling range restriction induce unacceptable accuracy loss and also cannot be applied MSDeformAttn dataflow directly.
% handle the attention workload via exploiting the $\mathbf{Q}\times\mathbf{K}^T$ sparsity xxxxxx xxxxxx xxxxxx xxxxxx xxxxxx xxxxxx xxxxxx xxxxxx xxxxxx xxxxxx xxxxxx xxxxxx xxxxxx.
% Deformable Attention reconstructs the attention operations with the grid-sample-based relation-graph building and parsing instead of regular matrix multiplication in traditional networks so that previous transformer-oriented DSAs are not suitable here. xxxxxx xxxxxx xxxxxx xxxxxx xxxxxx xxxxxx xxxxxx xxxxxx xxxxxx xxxxxx xxxxxx xxxxxx xxxxxx.
% Although some works \cite{huang2019algorithm, huang2021codenet, ahn2020efficient, li2022computational} also make great efforts on similar grid-sample-based operators like deformable convolution[cite Jifeng Dai Deformable Convolution], they attain unsatisfactory detection accuracy and lack of support for attention dataflow. xxxxxx xxxxxx xxxxxx xxxxxx xxxxxx xxxxxx xxxxxx xxxxxx xxxxxx xxxxxx xxxxxx xxxxxx xxxxxx.

Therefore, we propose \textbf{DEFA}, a dedicated MSDeformAttn accelerator with algorithm-architecture co-optimization.
The main contributions are as follows:
\begin{enumerate}
    \item We comprehensively characterize the performance bottlenecks of MSDeformAttn in deformable transformers and identify the root cause of deployment inefficiency. 
    \item At the algorithm level, we propose a pruning-assisted grid-sampling scheme by deeply exploiting the sampling redundancy.
    To reduce the memory access of sampled fmaps, frequency-weighted fmap pruning (FWP) is adopted to ignore 43\% unimportant pixels based on inter-layer pixel-wise acquisition.
    Meanwhile, probability-aware point pruning (PAP) is also applied to achieve an 84\% reduction in memory access through softmax-based sparsity exploitation.
    % Frequency-weighted fmap pruning (FWP) and probability-aware point pruning (PAP) are adopted to reduce memory requirement and access energy consumption respectively, 
    % we also present the softmax-based sampling point pruning (SoPr) method on grid-sample operations to reduce over 80\% computation costs and data access. 
    % To further simplify the sampling, the sampling range inside each scale feature map is limited. 
    \item At the hardware level, we design an efficient MSDeformAttn architecture while fully harnessing the performance gains from algorithm-level optimization. 
    we decouple the intra-level sampling computation to explore the multi-scale parallelism for bank conflict elimination, boosting the MSGS throughput by {3.06$\times$}.
    Furthermore, fine-grained operator fusion is adopted within each MSDeformAttn layer to avoid heavy data movement between sample and aggregation, which benefits from a reconfigurable PE array for MSGS and matrix computation.
    % Lastly, a reconfigurable PE array for the dual-mode MSDeformAttn procedure is developed to enhance the PE utilization rate and save computation resources. 
    \item Implemented on 40nm technology and extensively evaluated on representative benchmarks, MSDeformAttn achieves up to 10.1-31.9$\times$ speedup and 20.3-37.7$\times$ energy efficiency improvement over Nvidia RTX 2080Ti \& 3090Ti GPUs. Compared with the related accelerators, it improves the energy efficiency by 2.2-3.7$\times$, while supporting MSDeformAttn.
\end{enumerate}

%% file: 2_pre.tex
\section{Preliminary}

\subsection{Multi-Scale Deformable Attention}

MSDeformAttn identifies relations between each query and a small set of sampling points in multi-scale fmaps \(X \in \mathbb{R} ^{N_{in} \times D_{in} } \). 
Let  \(N_{in}\) and \(D_{in}\) denote the length of flattened feature maps from \(N_{l}\) levels \((N_{in} = \sum_{l=0}^{N_{l}-1} H_{l} \times W_{l})\) and hidden dimension of pixel vectors respectively. 
The computation of MSDeformAttn is shown as:
% \begin{equation}
% \begin{equation}
% \begin{split}
% MSDeformAttn(Q,p,X)&=Concat(head_{0},..., head_{H-1})\\
% \mathrm{where} \quad  head_{ij}&= Softmax(Q_{i}W_{j}^{A})  \cdot \left [(XW_{j}^{V})(p_i+\Delta p_j)\right] %\label{1}\\
% \end{split}
% \end{equation}
% \end{equation} 
% \[
% \mathrm{where}  \quad Head_{m}= \sum_{l=0}^{L-1} \sum_{k=0}^{K-1}A_{mlqk}\cdot W_{m}^{\prime} x^l(p_q+\Delta p_mlqk)
%  \]
% MSDeformAttn(z_{q},p{q},\left \{ x \right \}_{l=1}^L)=Concat(head_{0},..., Head_{M-1})
%head_{ij}= Softmax(Q_{i}W_{j}^{A})  \cdot V(p_i+\Delta p_j)
%X_{i}W_{i}^{\prime} 
%\sum_{l=0}^{N_{l}-1} \sum_{k=0}^{N_{p}-1}
%Head_{m}= \sum_{l=0}^{L-1} \sum_{k=0}^{K-1}A_{mlqk}\cdot W_{m}^{\prime} x^l(p_q+\Delta p_mlqk)
%\left \{ x \right \}_{l=1}^L where\mathrm{where} 
%\in \mathbb{R}^{N_{in}}
%\begin{equation}
%\begin{split}
%MSDeformAttn(Q,p,X)&=Concat(head_{0},..., head_{H-1})\\
%\mathrm{where} \quad  head_{ij}&= Softmax(Q_{i}W_{j}^{A})  \cdot V_{j}(p_i+\Delta p_j)  \label{1}\\
%  V&=XW^{V}
%\end{split}
%\end{equation}
%\begin{equation}
%\begin{split}
%MSDeformAttn(Q,p,X)&=Concat(head_{0},..., head_{H-1})\\
%\mathrm{where} \quad  head_{ij}&= Softmax(Q_{i}W_{j}^{A})   \\
%\end{split}
%\end{equation}
\begin{equation}\label{define}
\begin{split}
MSDeformAttn(\textbf{Q},\textbf{P},\textbf{X})&=Concat(\textbf{H}_{0},..., \textbf{H}_{N_{h}-1})\\
\mathrm{where} \quad  \textbf{H}_{ij}&= Softmax(\textbf{Q}_{i}\textbf{W}_{j}^{A})\textbf{V}_{j}(\textbf{P}_i+\Delta \textbf{P}_{ij}) \\
\textbf{V}=\textbf{X}\textbf{W}^{V},& \quad \Delta \textbf{P}=\textbf{Q}\textbf{W}^{S}
\end{split}
\end{equation}
where \(\textbf{Q}\in \mathbb{R} ^{N_{in} \times D_{in} } \) and \(\textbf{V}\in \mathbb{R} ^{N_{in} \times D_{in} } \) denote the query matrix and the multi-scale fmaps, 
\(i\) indexes the row vector in the matrix and \(j\) indexes the \(N_{h}\) heads. 
\(\textbf{W}^A\in \mathbb{R} ^{D_{in} \times D_{in} } \), \(\textbf{W}^V\in \mathbb{R} ^{D_{in} \times D_{in} } \) and \(\textbf{W}^S\in \mathbb{R} ^{D_{in} \times (2N_{h}N_{l}N_{p})} \) are all learnable weights, where \(N_{p}\) denotes the fixed number of sampling points in each level of fmaps. 
\(\textbf{H}_{ij}\in \mathbb{R} ^{N_{in} \times D_{h} }\) (\(D_{h}=D_{in}/N_{h}\))  is the \(i^{th}\) row vectors of the output matrix in the \(j^{th}\) attention head. 
\(\textbf{P}_i\in \mathbb{R} ^{N_{in} \times N_{l} \times 2} \) enumerates \(N_{in}\) coordinates of regular grids over each level fmap.

\begin{figure}[t]
  \centering
  \includegraphics[width=\linewidth]{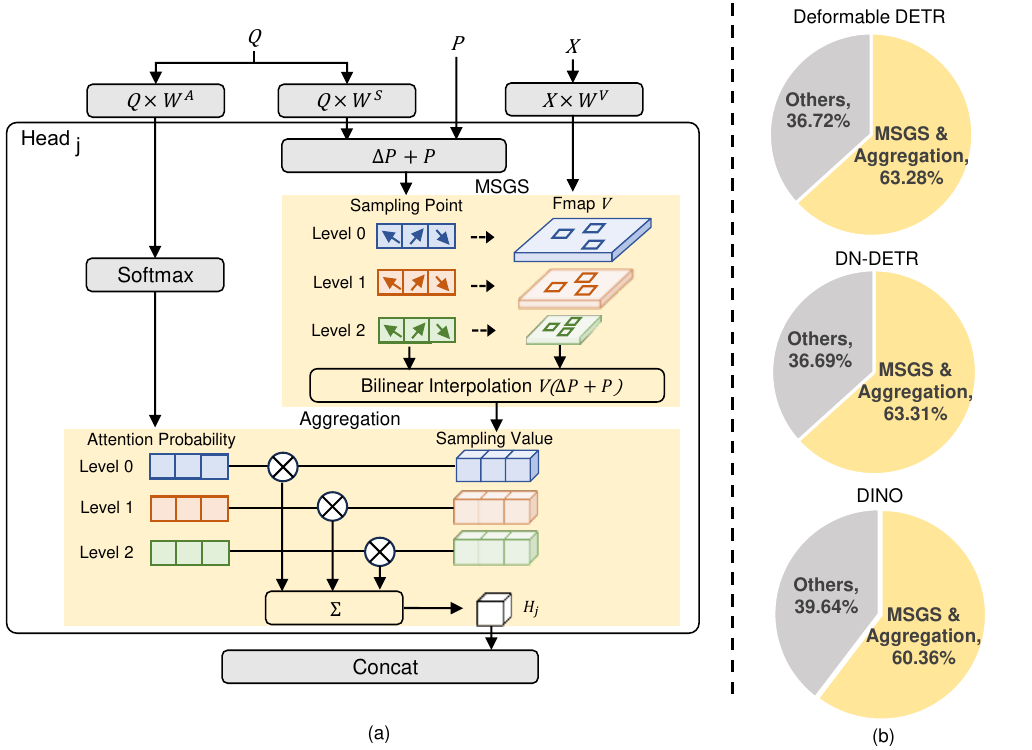}
  \caption{(a) Introduction of MSDeformAttn. (b) MSDeformAttn latency breakdown.}
  \label{fig:2_deformattn}
\end{figure}

Figure \ref{fig:2_deformattn} (a) illustrates the whole procedure of MSDeformAttn.
In each head of MSDeformAttn, the softmax unit normalizes all points in different levels projected from a row vector of the input query and generates the attention probability vector.
The MSGS procedure adopts the bilinear interpolation (BI) kernel to process the fractional sampling points and obtains sampling value from the multi-scale fmaps.
In the aggregation stage, each pixel vector of the sampling value is multiplied by an element of the attention probability vector and then all the weighted vectors are summed to gain an output vector of a head.

\subsection{Computational Properties Analysis}

% \textcolor{red}{
% 1. Analyze the runtime breakdown, sampled fmap size and sampling point number in DeformAttn.
% 2. Analyze the workload increase compared to deformable convolution, especially the sampled fmap size and sampling point number.
% 3. Analyze why the previous attention accelerators cannot efficient support MSDeformAttn.
% }
As shown in Figure \ref{fig:2_deformattn} (b), We profile the MSDeformAttn latency breakdown on Deformable DETR \cite{zhu2021deformable}, DN-DETR \cite{dndetr} and DINO \cite{dino} on Nvidia RTX 3090Ti.
MSGS and aggregation account for over 60\% of the inference runtime of MSDeformAttn, while their computation cost only takes 3.25\%, demonstrating severe inefficiency on GPU platforms. 
This is mainly caused by irregular memory access in MSGS.
In MSGS, unbounded sampling coordinates are dynamically generated and incur unpredictable memory access. 
It violates the locality principles commonly used to reduce DRAM access, resulting in minimal fmap reuse.
Moreover, as multi-scale fmaps in MSGS have more pixels than single-scale fmaps, the sampling points are dispersed over a wider range, which further increases memory access.
In addition, the number of sampling points in MSGS also grows proportionally with the number of fmap pixels, aggravating the inefficiency of the unordered sampling process.

Although DeformConv also uses a similar grid-sampling module, the workload of MSGS in MSDeformAttn is multiple times higher than DeformConv.
% The multi-scale fmaps in MSDeformAttn are 21.3\(\times\) larger than the single-scale fmaps in DeformConv, entailing extremely larger on-chip buffer requirement.
% Besides, the sampling points in MSDeformAttn are also \(N_{l}N_{p}\times\) more than in DeformConv in each head, which results in more irregular memory access in MSGS.
% Hence, the 
In particular, the multi-scale fmaps in MSDeformAttn are 21.3\(\times\) larger than the single-scale fmaps in DeformConv.
Additionally, the sampling points in MSDeformAttn are also \(N_{l}N_{p}\times\) more than those in DeformConv in each head.
This causes an extremely larger on-chip buffer requirement and more irregular memory access in MSGS.
% , thus leading to higher inefficiency for MSDeformAttn process.

Existing attention accelerators \cite{elsa, spattn2021, qinghua2023} reduce the computation of weak-related tokens via random projection (ELSA \cite{elsa}), attention score sort (SpAtten \cite{spattn2021}) and approximate computation (BESAPU \cite{qinghua2023}).
However, MSDeformAttn does not explicitly compute token relevance, making their methods inapplicable.
% However, there is no explicit computation of token relevance in MSDeformAttn, so their methods are inapplicable to MSDeformAttn.
% devises the dynamic pruning method to detect wea exploit sparsity in attention to optimize computation.
% ELSA \cite{elsa} estimate the similarity between queries and keys with sign random projection and prune irrelevant pairs.
% Spatten \cite{spattn2021} performs cascade pruning on the attention tokens and heads.
% \cite{qinghua2023} predicts weak-related tokens with the local property of attention and approximately compute them. 
Besides, they do not efficiently support the MSGS procedure and lack optimization on the dataflow of MSGS.
In MSGS, irregular memory access on multi-scale fmaps requires attention accelerators with up to 9.8MB on-chip buffer size, significantly increasing area and reducing efficiency.
% In MSGS, the irregular memory access on multi-scale fmaps demands the attention accelerators for up to 9.8MB on-chip buffer size, which significantly enlarges the area and damages efficiency. 
DEFA is the first work to exploit sparsity in fmaps and sampling points and to explore the multi-scale parallelism in MSGS processing. 

\begin{figure}[t]
  \centering
  \includegraphics[width=\linewidth]{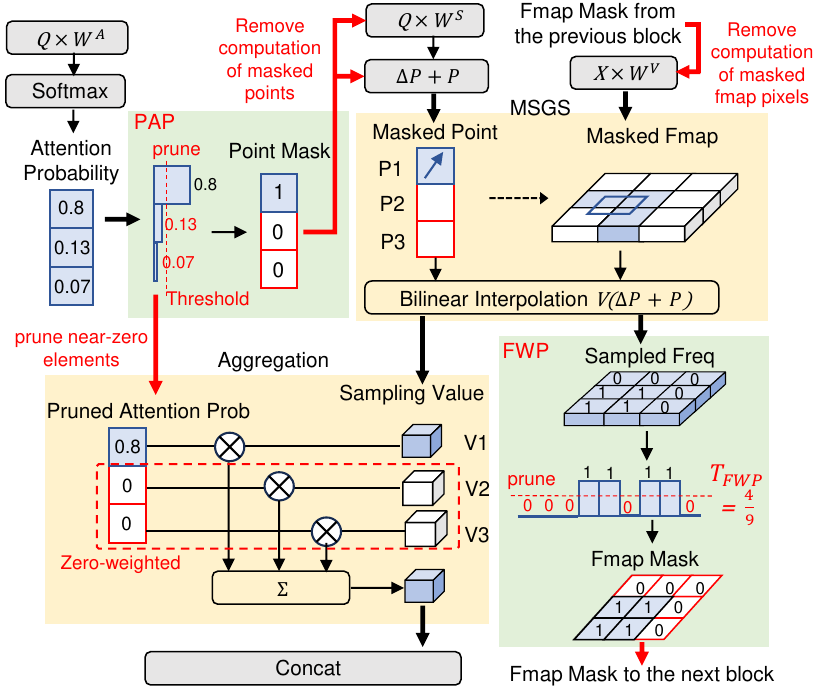}
  \caption{Overview of sparsity-aware grid-sampling.}
  \label{fig:3_grid_sample}
\end{figure}

% \emph{Shortcomings of related works and the reasons why they cannot suit the MSDeformAttn}.
% Many attention accelerators exploit sparsity in softmax results and diminish the calculation of these zero results.
% xxxx xxxx xxxx xxxx xxxx xxxx xxxx xxxx xxxx xxxx xxxx xxxx xxxx xxxx xxxx xxxx xxxx xxxx xxxx xxxx xxxx.
% The prune strategies and dedicated hardware designs in these works leverage specific execution pattern \(\textbf{QK}^{T}\) in self-attention, but MSDeformAttn multiplies \(\textbf{Q}\) with learned weights and unfits their sparsity prediction algorithm. 
% Apart from that, these hardware designs also require corresponding modules to deploy Grid-Sampling in MSDeformAttn.
% There are prior DSAs proposed to enhance the efficiency of Grid-Sampling in deformable convolution as well.
% xxxx xxxx xxxx xxxx xxxx xxxx xxxx xxxx xxxx xxxx xxxx xxxx xxxx xxxx xxxx xxxx xxxx xxxx xxxx xxxx xxxx.
% However, deformable convolution is a degenerated variant of MSDeformAttn, attending to only one sampling point in the input single-scale feature map for every attention head. With more sampling points distributed over multi-scale feature maps, Grid-Sampling of MSDeformAttn imposes higher on-chip buffer size and parallelism requirement on hardware, which is impossible for current DSAs to satisfy.
% Besides, in these works, the aggressive restriction on the sampling field is inflexible for multi-scale features and incurs an unacceptable accuracy drop in MSDeformAttn. 

%% file: 3_algo.tex
\section{Sparsity-Aware Grid-Sampling for MSDeformAttn Dataflow}

% \emph{Short Introduction of your motivation and proposed approaches.}
% In MSGS, numerous unnecessary pixels exist in the multi-scale fmaps and 
In MSGS, numerous fmap pixels and sampling points cause enormous memory access and have high redundancy.
To remove unnecessary memory access and enhance the efficiency of MSGS, we exploit the sparsity in multi-scale fmaps and sampling points via FWP and PAP approaches respectively. 
% With the purpose of alleviating on-chip storage requirements, we also propose the level-wise range narrowing method for multi-scale feature maps.  
%\subsection{Probability-aware Pruning}
\subsection{Frequency-Weighted Fmap Pruning}

% \emph{To restrict the size of sampled images, we analyze the probability of each pixel in multi-scale images on the fly and remove the least probably sampled pixels}.
% We introduce the sampled frequency of each pixel in the multi-scale fmaps and propose FWP that restricts the size of large fmaps.
To restrict the size of the large sampled fmaps, we analyze the sampled frequency of each pixel on them and find that it shows a non-uniform distribution.
This demonstrates that a small proportion of pixels has a much higher probability of being accessed and is essential to the detection accuracy.
FWP is proposed to remove other unimportant pixels with low sampled frequency.
Firstly, the sampled times of every pixel in the fmap are counted in MSGS of the MSDeformAttn block and indicate the sampled frequency.
As shown in Figure \ref{fig:3_grid_sample} right, the neighboring points of the sampling point in BI are accessed once, so the sampled frequency of these pixels is counted to 1 and the sampled frequency of the others is 0. 
Then, the pixels with a lower sampled frequency than the preset threshold are pruned and their locations are recorded in a bit mask as the fmap mask.
Finally, the generated fmap mask is applied in the next MSDeformAttn block to eliminate the linear projection and memory access of the masked fmap pixels.
Here the threshold in FWP is defined as:
\begin{equation}\label{threshold}
\begin{split}
T_{FWP} = k\cdot \frac{1}{H W}\sum_{i=0}^{H W-1} F_{i}
\end{split}
\end{equation}
where \(k\) is a hyperparameter and \(F_{i}\)\ denotes the sampled frequency of the \(i^{th}\)\ pixel in the fmap of size \({H W}\). 
In Figure \ref{fig:3_grid_sample} right, \(k\) is assumed as 1 and the fmap size is 9.
We adjust \(k\) to achieve a trade-off of accuracy and sparsity in the finetuning.
\subsection{Probability-Aware Point Pruning}

% \emph{To reduce the number of sample points, we utilize the sparsity from Softmax and propose softmax-based pruning}.
% Facing another challenge of numerous sampling points,
We propose PAP to detect and remove unessential sampling points in MSGS using their related attention probabilities as illustrated in Figure \ref{fig:3_grid_sample} left.
% that explore the sparsity in the normalized attention probability after softmax. 
% The MSGS generates a sampling value
In the aggregation process, the sampling values in an attention head are multiplied by the normalized attention probabilities from softmax and then summed up. 
The summation of attention probabilities is confined to 1 and their differences are exponentially amplified.
We set a threshold to filter out the near-zero attention probabilities and they constituted a dominant proportion (over 80\% in Deformable DETR).
The sampling values weighted by them make small contributions to the results of aggregation and detection accuracy. 
Hence, the near-zero attention probabilities are pruned and the point mask is established as a bit mask to eliminate the following processing of the sampling points generating the zero-weighted sampling values in the current MSDeformAttn block.
In Figure \ref{fig:3_grid_sample} left, the zero-weighted sampling values \(V2, V3\) are multiplied with 0 in the aggregation process after PAP, and thereby the sampling points \(P2, P3\) producing \(V2, V3\) are unnecessary and removed to save computation cost and memory access.
\begin{figure}[t]
  \centering
  \includegraphics[width=\linewidth]{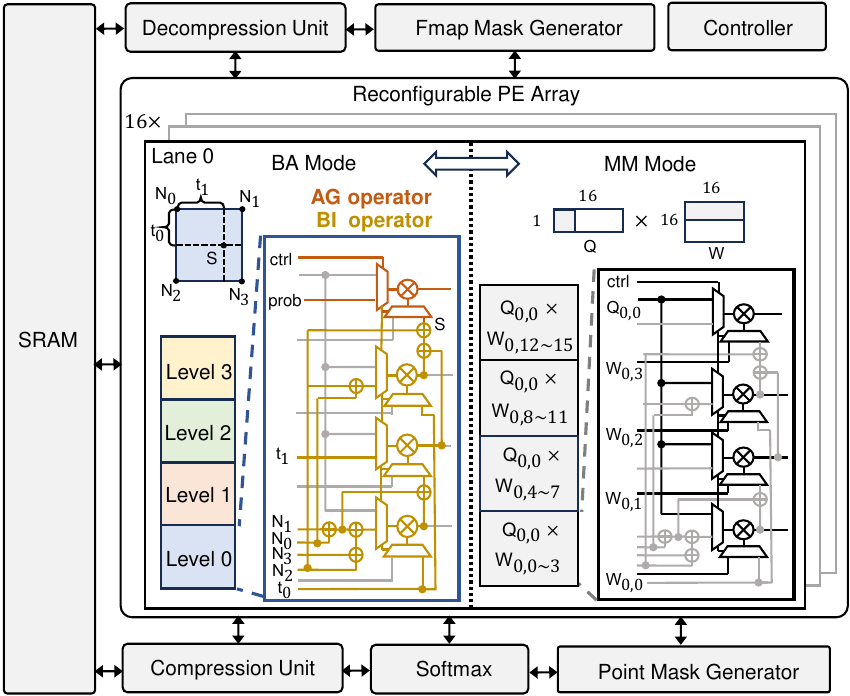}
  \caption{Overview of DEFA architecture.}
  \label{fig:4_arch_overview}
\end{figure}

% \emph{We still need to store huge sizes of images on-chip and the random and irregular sampling within each image still cannot be avoided.
% Therefore, the range of sampling is further constrained while training}.
% To overcome the difficulty in unfeasible on-chip storage of multi-scale features in MSDeformAttn Grid-Sampling, we constrain the sampling offset range around every reference point and fine-tune the model to recover the accuracy loss. 
% Through this method, processing all sampling points around a reference point only requires much less on-chip buffer size to store the pixels in the bounded range.
% In addition, we observe that imposing unified restrictions on each level of multi-scale feature maps causes distinct accuracy loss.
% Based on the observation, a level-wise range narrowing strategy is developed to bounded multi-scale feature maps in different ranges. 
% As shown in Figure \ref{fig:3_grid_sample} (b), lower-level feature maps with the larger size are less sensitive to the range restriction.
% Compared to the original bounded method marked in the dotted red box, this flexible approach marked in the solid red box can further decrease on-chip memory storage by 25\% without extra influence on the accuracy.

%% file: 4_arch.tex
\section{High-throughput Deformable Attention Architecture}

% \emph{Short Introduction of your proposed architecture, especially the function of each part.} 
In this section, we present DEFA architecture co-designed with the pruning algorithms to efficiently process MSDeformAttn in Figure \ref{fig:4_arch_overview}. 
The fmap mask generator and sampling point mask generator implement FWP and PAP, respectively. 
The compression unit decompression unit eliminates the redundant bandwidth and computation of the masked data. 
% To support the parallel processing of MSGS, the multi-scale fmaps arrangement in the SRAM is developed for no conflict. 
The reconfigurable PE array can switch between the matrix multiplication (MM) mode and the BI mode to accelerate MM or fine-grained operator fusion.
% enhances not only the throughput of bilinear interpolation and aggregation but also the utilization rate of multipliers.
% Section 4.1 introduces the three main dataflows of MSDeformAttn. Specifically, Section 4.2 explains the multi-scale parallel processing in Grid-Sampling. At last, we clarify the two working modes of configurable PE array in Section 4.3.

\begin{figure}[t]
  \centering
  \includegraphics[width=\linewidth]{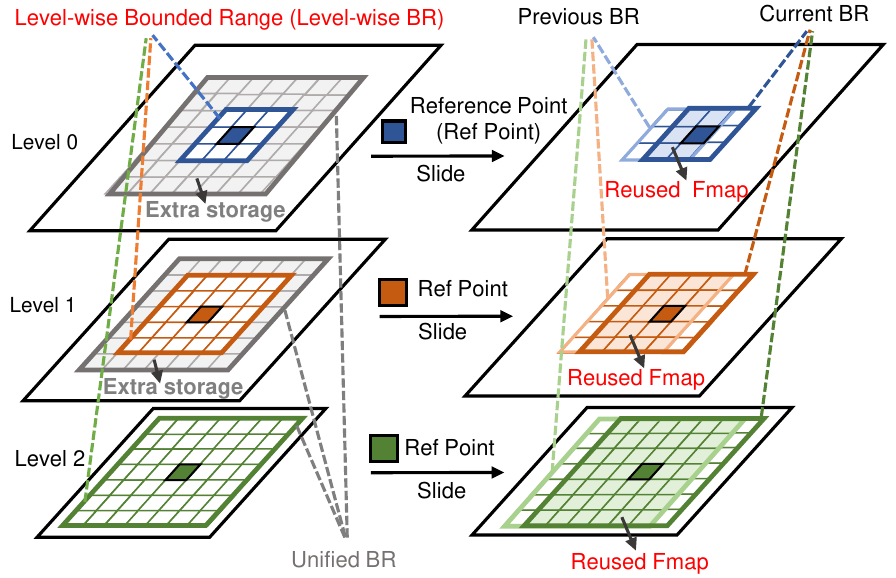}
  \caption{Level-wise range-narrowing scheme.}
  \label{fig:5_reuse}
\end{figure}

\subsection{Dataflow Overview}

% \emph{Introduce your dataflow mapping method, especially the fine-grained layer fusion and ref range reuse}.
 % shows an overview of DEFA architecture.
% DEFA rearranges the operators in MSDeformAttn so that the proposed pruning approaches can be applied to reduce computation and memory access. 
DEFA rearranges the operators in MSDeformAttn to enable the application of FWP and PAP, reducing computation and memory access.
Firstly, attention probabilities are calculated and the point mask is updated. 
Then, the masked sampling points pruned by the point mask are generated in the reconfigurable PE array in the MM mode, which includes \(\Delta \textbf{P}=\textbf{Q}\textbf{W}^{S}\) in Eq.\ref{define}.
Next, the linear projection of the masked fmaps (\(\textbf{V}=\textbf{X}\textbf{W}^{V}\) in Eq.\ref{define}) is pruned by the fmap mask from the last MSDeformAttn block and performed in the PE array.
% the reconfigurable PE array in the MM mode generates the sampling points and the fmaps pruned by the point mask and the fmap mask respectively. 
Finally, DEFA processes the fused MSGS and aggregation operators with the reconfigurable PE array in the BA mode.
Meanwhile, the fmap mask generator receives the sampling address in the BI and executes FWP for the next block.
% Finally, DEFA processes the fused MSGS and aggregation operators with the reconfigurable PE array at the BA mode.
% The reconfigurable PE array at the BA mode processes the fused MSGS and aggregation operators with 
% Meanwhile, the fmap mask generator receives the sampling address in the bilinear interpolation and runs FWP for the next block.
% Then, the sampling offsets pruned by the point mask are computed in the reconfigurable PE array at the MM mode.
% The masked fmap generation is performed in the PE array.

Inspired by the prior work on DeformConv \cite{huang2021codenet}, we propose the level-wise range-narrowing scheme to decrease on-chip storage in MSGS, as illustrated in Figure \ref{fig:5_reuse} left.
DEFA uses bounded ranges of different sizes to limit the sampling offsets around a reference point based on the level of multi-scale fmaps. This is because the bounded range in some levels can be further reduced without any accuracy loss.
% DEFA adopts different sizes of bounded ranges to restrict the sampling offsets around a reference point according to the level of multi-scale fmaps, which is based on that the bounded range in some levels can be further shrunk without increment of accuracy loss.
% When processing the sampling points extended from a ref point, DEFA fetches the pixels in the bounded range from off-chip storage.
Applying unified restriction on all levels of the multi-scale fmaps causes an extra 25\% storage requirement. 
Figure \ref{fig:5_reuse} right shows that when the reference point slides to the next pixel in fmaps,
the overlapping pixels between the previous and the current bounded range are reused to avoid repetitive memory access.

\begin{figure}[t]
  \centering
  \includegraphics[width=\linewidth]{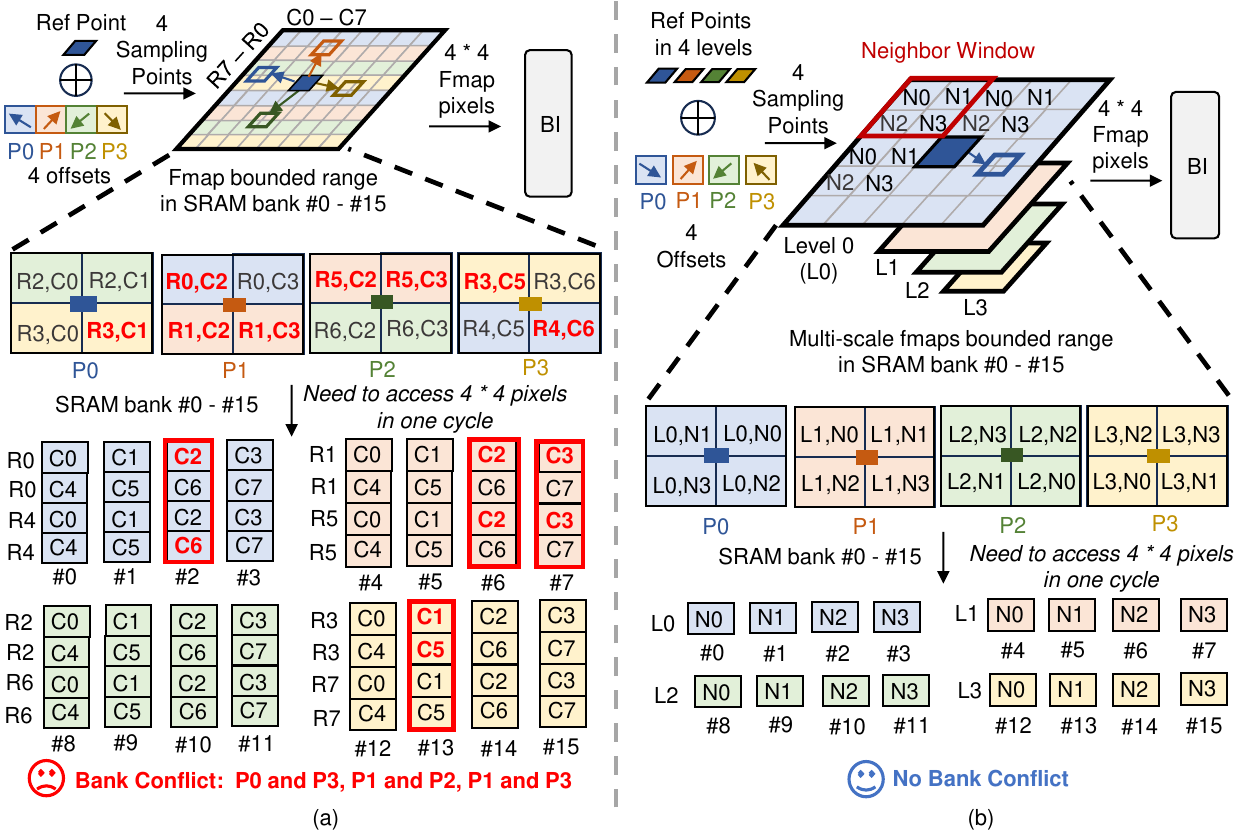}
  \caption{Illustration of (a) intra-level parallel processing and (b) inter-level parallel processing.}
  \label{fig:4_parallel}
\end{figure}

\subsection{Multi-Scale Parallel Processing}

% \emph{Introduce the multi-scale parallel processing }.
To minimize latency and optimize hardware resource utilization, we investigate both intra-level and inter-level parallel processing in MSGS. 
% Figure \ref{fig:4_parallel} illustrates the multi-scale parallel processing of MSGS.
% In the BI of MSGS, DEFA aims to calculate 4 sampling points in parallel without blocking, so 16 fmap pixels (4 neighboring pixels for each sampling points) need to be accessed from 16 SRAM banks in one cycle.
The purpose of DEFA in the BI of MSGS is to calculate four sampling points in parallel without bank conflicts. 
This requires accessing 16 fmap pixels, with four neighboring pixels for each sampling point, from 16 SRAM banks in one cycle.
If DEFA processes four sampling points in one level of multi-scale fmaps, which is intra-level parallel processing presented in Figure \ref{fig:4_parallel}(a), all the pixels within the bounded range centered on the reference point are stored in the SRAM. 
This can cause bank conflicts when multiple sampled pixels are stored in the same bank.
To prevent this issue, we propose inter-level parallel processing based on the observation that sampling points are only located in the same level of multi-scale fmaps as their reference points, which is utilized to limit the sampling range in their level. 
% In the inter-level parallel processing, DEFA processes sampling points in different levels with the proposed memory arrangement, 
As illustrated in Figure \ref{fig:4_parallel}(b), compared to intra-level parallel processing, DEFA processes four sampling points extended from the reference points in four levels of multi-scale fmaps.
In addition, the pixels within the bounded range of every level are stored in every four banks of the overall 16 SRAM banks.
The bounded range is tiled into several \textit{Neighbor Windows}, and the four pixels in each \textit{Neighbor Window} are mapped to the four SRAM banks.
This approach allows for the sampled pixels to be accessed in different banks without any bank conflicts.

\subsection{Fine-Grained Operator Fusion with the Reconfigurable PE Array}

% We propose fine-grained operator fusion of MSGS and aggregation with a reconfigurable PE array.
% DEFA directly multiplies the sampling values from BI with the attention probability stored in the additional buffer so that the weighted sum in the aggregation is obtained without external memory access to the sampling values.
% To support the fused operators, the reconfigurable PE array is designed to be configured in either BA mode or MM mode as shown in Figure \ref{fig:4_arch_overview}.
% In the BA mode, the PE array processes sampled pixels from 4 levels in parallel, and each lane in the array is assigned one of the 16-pixel channels.     
% At the MM mode, the PE array computes the MM between a 16-element row vector and a 16 \(\times\) 16 tile in the output-stationary dataflow. 

We propose fine-grained operator fusion of MSGS and aggregation, removing the off-chip transfer of sampling value.
To support the process with the limited computing resources in the PE array, we transform BI and design a reconfigurable PE array alternating between the BA mode and the MM mode, as shown in Figure \ref{fig:4_arch_overview}.
Assume there is a sampling point, denoted as \(S\) at \((x, y)\), with four neighboring points denoted as \(N_{0}\) at \((x_{0}, y_{0})\) (top left), \(N_{1}\) at \((x_{1}, y_{0})\) (top right), \(N_{2}\) at \((x_{0}, y_{1})\) (bottom left) and \(N_{3}\) at \((x_{1}, y_{1})\) (bottom right) respectively. Then BI is implemented as:
\begin{equation}\label{bi}
\begin{split}
S = N_{0}(x_{1} - x)(y_{1} - y) + N_{1}(x - x_{0})(y_{1} - y) \\
+ N_{2}(x_{1} - x)(y - y_{0}) + N_{3}(x -x_{0})(y - y_{0}) 
\end{split}
\end{equation}
As the coordinates of the four neighboring points are all integers and located at the four corners of the grid, \(x_{1}\) and \(y_{1}\) are equal to \(x_{0}+1\) and \(y_{0}+1\) respectively. After replacing them and a series of transformations, Eq.\ref{bi} becomes:
% S = N0 + (N2 - N0) * t0 + {(N1 - N0) + [N3 - N2 - (N1 - N0)] * t0} * t1				(2)
\begin{equation}\label{bi_change}
\begin{split}
S = N_{0} + (N_{2} - N_{0})t_{0} 
+ [ (N_{1} - N_{0}) \\
+ (N_{3} - N_{2} - N_{1} + N_{0})t_{0} ] t_{1}
\end{split}
\end{equation}
Where \(t_{0}\) = \(y\) - \(y_{0}\) and \(t_{1}\) = \(x\) - \(x_{0}\). The calculation of \(t_{0}\) and \(t_{1}\) is performed in other units. As a result, the BI operator part in Figure \ref{fig:4_arch_overview} only employs three multipliers and seven adders. The AG operator part performs the multiplication of attention probability and \(S\).
Additionally, the PE array in the MM mode conducts MM between a 16-element vector (\(Q\)) and a 16 \(\times\) 16 tile (\(W\)) in the output-stationary dataflow.

%% file: 5_eval.tex
\section{Evaluation}

\subsection{Experimental Methodology}

\subsubsection{Benchmarks}
% \emph{Introduce your networks and datasets. Emphasize how you implement the inference flow on GPUs like using the official repository and so on}.
We evaluate DEFA on MSDeformAttn layers in the encoders of Deformable DETR (De DETR) \cite{zhu2021deformable}, DN-DETR  \cite{dndetr} and DINO \cite{dino}.
The evaluation task is object detection on the COCO 2017 dataset \cite{coco}.
We utilize PyTorch to conduct experiments on the benchmarks with reference to the official implementation. 
Finetuning is applied to the modified models by the software methods in Section 3 to recover the accuracy.
% frequency-aware pruning, softmax-based pruning and level-wise range-narrowing to recover the accuracy.
The MSDeformAttn modules in the encoder layers of the models are quantized to 12bits during the inference. 
We compare DEFA with NVIDIA RTX 2080Ti and 3090Ti GPUs and SOTA attention accelerators \cite{elsa, spattn2021, qinghua2023}. 
% The following exhibits the detailed discussions.
% We reproduce the benchmarks with PyTorch referring to the official implementation [46]. Based on that, the dataflow-decoupling GFLNs are implemented for accuracy and performance verification, which are quantized to 8bits and retrained to recover the accuracy.

\subsubsection{Hardware Implementation}
DEFA is described in SystemVerilog and synthesized with Synopsys Design Compiler for 400MHz clock frequency to estimate the area and power under a 40nm technology. 
We implement a cycle-accurate simulator to model the computation and memory access and evaluate the performance of DEFA.
We obtain the area and energy consumption of SRAM with CACTI \cite{cacti}, and a moderate 256GB/s HBM2 is used as the external memory system, consuming 1.2pJ/b \cite{dram} for data access.
% , which consumes 1.2pJ/b \cite{dram} for data access.  

% \emph{My version on TVLSI:}
% We implement a prototype of FLNA with SystemVerilog and synthesize it with Synopsys Design Compiler under a 40nm CMOS technology. A cycle-accurate simulator referring to [41] is developed to model the behavior of computing and memory access and investigate the precise performance of FLNA. Regarding the memory, on-chip SRAM area and energy are obtained with CACTI [42], and we use a moderate HBM2 with 256GB/s bandwidth as the external memory system, which consumes 1.2pJ/b [43] for data access.

% \subsubsection{Baselines}
% We compare DeformAcc with NVIDIA RTX 2080Ti and 3090Ti GPUs, Intel Xeon Gold 5220R CPU @2.20GHz, SOTA deformable convolution accelerators and attention accelerators \cite{}, \cite{}. The following exhibits the detailed discussions.

% \emph{My version on TVLSI:}
% Various hardware platforms are selected as the performance baselines, such as NVIDIA RTX 2080Ti [44] and 3090Ti [45] GPUs, Intel Xeon Gold 5220R CPU @2.20GHz, and SOTA point-cloud-based accelerators [29], [30], [33].  Detailed discussions are exhibited in the following.

\subsection{Algorithm Evaluation}

\begin{figure}[t]
  \centering
  \includegraphics[width=\linewidth]{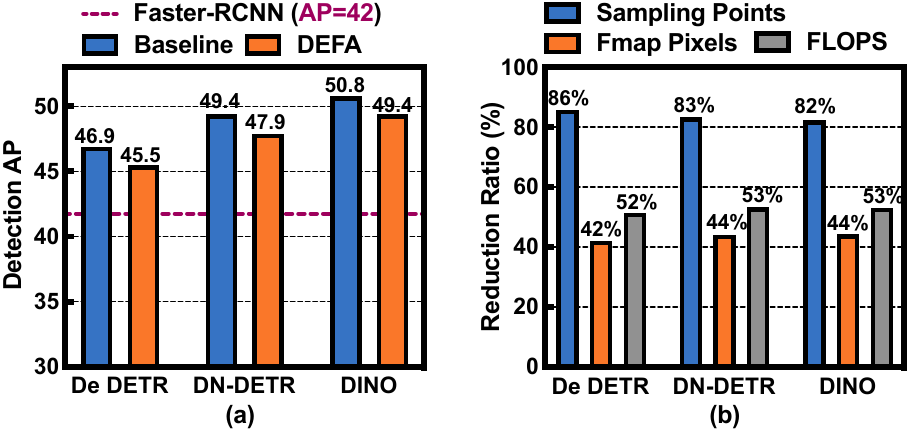}
  \caption{(a) Detection Average Precision of our methods and comparisons with other works. (b) Reduction in sampling points, fmap pixels, and computation cost.}
  \label{fig:5_algorithm_accuracy}
\end{figure}

Figure \ref{fig:5_algorithm_accuracy} (a) presents the standard average precision (AP) of our methods and comparisons with the baseline of the benchmarks as well as the Faster R-CNN \cite{FasterR-CNN}. 
The processing of FWP, PAP, level-wise range-narrowing, and INT12 quantization causes 0.8, 0.3, 0.26, and 0.07 AP drop on average on the benchmarks, respectively.
Compared to INT12 quantization, INT8 quantization is not adopted because it results in an average drop of 9.7 AP on the benchmarks, which is an unacceptable accuracy degradation.
DEFA preserves a relatively high detection accuracy with negligible AP loss, which is 3.5-7.4 AP higher than Faster R-CNN.
Figure \ref{fig:5_algorithm_accuracy} (b) shows the reduction ratio in sampling points, fmap pixels, and computation cost achieved by our pruning algorithms. 
% Figure \ref{fig:5_algorithm_accuracy} (b) shows the average reduction of fmap pixels, sampling points and computation in the tested benchmarks as well as the accurate AP impact analysis.
FWP and PAP reduce 43\% fmap pixels and 84\% sampling points on average, and the computation cost on the unimportant fmap pixels and sampling points is also eliminated, accounting for more than 50\% of the overall computation.
% The acceptable AP degradation caused by the two pruning methods demonstrates their effectiveness. 
For level-wise range-narrowing, we adjust bounded ranges of sampling offsets in each level to achieve a trade-off between accuracy and SRAM size. 
% After finetuning, all benchmarks present an accuracy loss of less than 2 mAP and the MSDeformAttn layers in the encoders of the models obtain \% GFLOPs reduction.
% With the support of sparse computation, our co-designed DeformAcc can utilize the decrease in computation cost to effectively accelerate the sparse MSDeformAttn with an acceptable accuracy loss.

% Figure \ref{fig:5_algorithm_accuracy} (a) presents the impact of the pruning and range-narrowing scheme on mean Average Precision (mAP) for three benchmarks, while Figure \ref{fig:5_algorithm_accuracy} (b) shows the reduction in computation cost between the base and pruned MSDeformAttn.
% For frequency-aware pruning and softmax-based pruning, we set the thresholds to achieve a sparse rate of 40 \% to 50 \%  in sampling feature maps and a sparse rate of 80 \% to 90 \% in sampling points respectively.
% For level-wise range-narrowing, we adjust bounded ranges of sampling offsets in each level to realize a trade-off between accuracy and SRAM size. 
% After sparsification and finetuning, all benchmarks present an acceptable accuracy loss of less than 2 mAP, but the MSDeformAttn.

\subsection{Performance Gain from Our Hardware Optimization Tactics}

\begin{figure}[t]
  \centering
  \includegraphics[width=\linewidth]{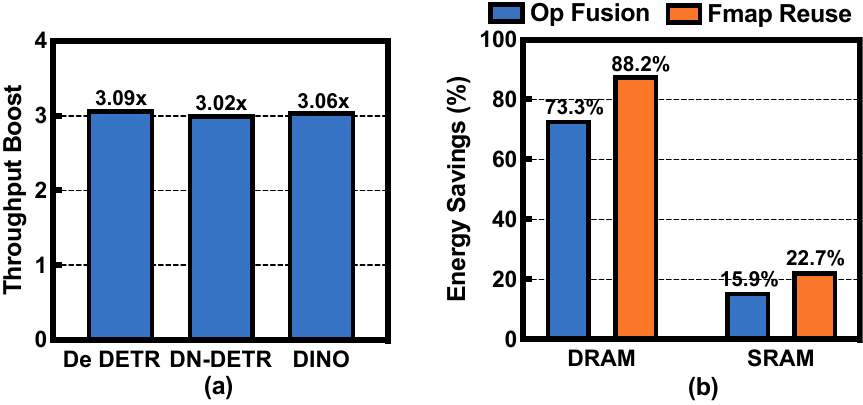}
  \caption{(a) MSGS throughput boost of inter-level parallel processing over intra-level parallel processing. (b) Energy savings of fine-grained operator fusion and fmap reuse.}
  \label{fig:5_optimization}
\end{figure}

\subsubsection{Multi-scale Parallel Processing}
% Compared to intra-level parallel processing, inter-scale parallel processing attains the same degree of parallelism but completely eliminates bank conflicts.  
Figure \ref{fig:5_optimization} (a) indicates the MSGS throughput improvement of inter-level parallel processing over intra-level parallel processing on the selected benchmarks.
The bank conflicts are detected in intra-level parallel processing when plural sampling points need to access different addresses in the same SRAM bank in one clock cycle. 
In this case, extra clock cycles are spent on detecting bank conflicts, stopping the pipeline, and sequentially processing the requests.
The proposed inter-level parallel processing completely avoids bank conflicts and achieves 3.06\(\times\) higher MSGS throughput than the intra-level parallel processing on average under the same degree of parallelism.

\subsubsection{Fine-grained Operator Fusion}
As shown in Figure \ref{fig:5_optimization} (b), fine-grained operator fusion (op fusion) of MSGS and aggregation results in 73.3\% and 15.9\% energy saving on DRAM access and SRAM access of the overall MSGS energy consumption in memory access.
Fine-grained operator fusion reduces the off-chip transfer of the BI result and utilizes the BI result directly to compute aggregation in the reconfigurable PE array in the BA mode, avoiding SRAM access.
DEFA only adds 0.5\% extra SRAM storage to support fine-grained operator fusion. 

\subsubsection{Fmap Reuse}
Figure \ref{fig:5_optimization} (b) presents the energy saving of fmap reuse.
Fmap reuse significantly reduces DRAM access of the fmap pixels in the overlapping bounded range, saving 88.2\% of the total MSGS energy consumption in memory access.
The writing operations to the SRAM of the repetitive fmap pixels fetched from the DRAM are also eliminated, which reduces 22.7\% energy consumption of the overall MSGS energy consumption in memory access.

\subsection{Comparisons with Other Platforms}

\begin{figure}[t]
  \centering
  \includegraphics[width=0.9\linewidth]{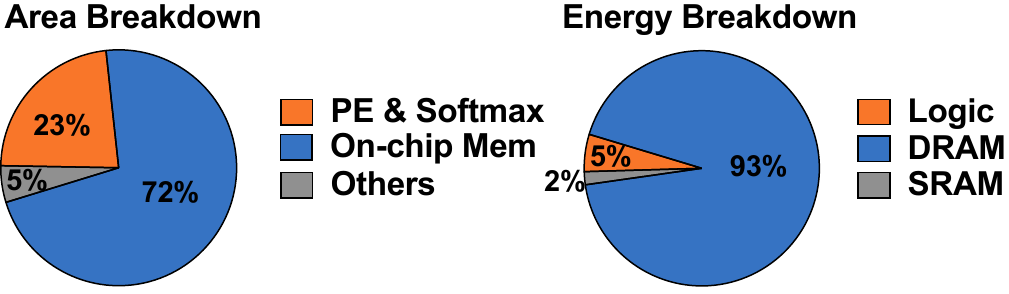}
  \caption{Area breakdown and energy breakdown of DEFA.}
  \label{fig:5_breakdown}
\end{figure}

\begin{figure}[t]
  \centering
  \includegraphics[width=\linewidth]{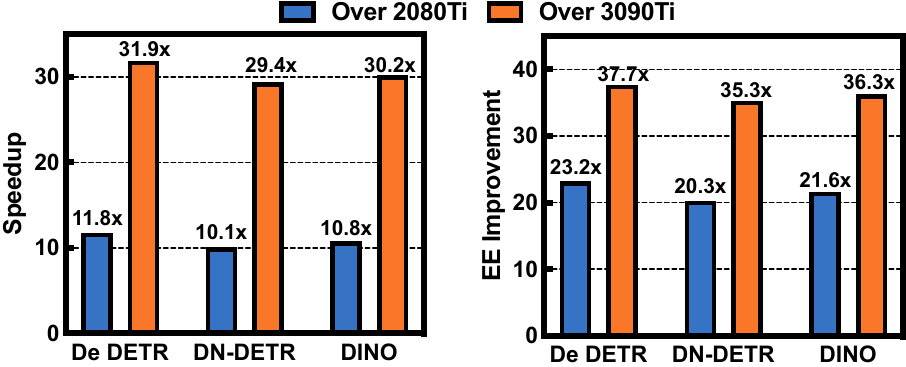}
  \caption{Speedup and energy Efficiency improvement of DEFA over GPUs.}
  \label{fig:5_comparison}
\end{figure}

We compare the speedup and the energy efficiency (EE) improvement of DEFA with the Nvidia GPUs in Figure \ref{fig:5_comparison}. 
The performances of CPUs are not evaluated because MSDeformAttn only has a CUDA implementation.
During the comparison with GPUs, we scale up DEFA to attain 13.3 TOPS and 40 TOPS peak throughput respectively, which matches Nvidia 2080Ti (13.5 TFLOPS\(@\)FP32) GPU and Nvidia 3090Ti (40 TFLOPS\(@\)FP32) GPU.
DEFA achieves 10.1-31.9$\times$ speedup as well as 20.3-37.7$\times$ energy efficiency improvement over Nvidia 2080Ti GPU (250W) \& Nvidia 3090Ti GPU (450W) on average.
The high speedup results from the inter-level parallel processing without any bank conflicts in MSGS.
DEFA further accelerates MSDeformAttn via FWP and PAP reducing the computation of redundant fmap pixels and sampling points.
The energy saving is attributed to DRAM and SRAM access reduction through fine-grained operator fusion and fmap 
reuse.
Figure \ref{fig:5_breakdown} shows the area and energy breakdown of DEFA. The SRAM occupies the largest area since MSGS requires a large amount of on-chip memory to store multi-level fmaps. 
The DRAM access dominates the energy consumption due to the large data transfer in MM.  

Table \ref{tab:V_Performance_Comparison} compares DEFA and other attention asic platforms \cite{elsa, spattn2021, qinghua2023}.
DEFA achieves 3.7\(\times\), 3.4\(\times\) and 2.2\(\times\) higher energy efficiency than ELSA \cite{elsa}, SpAtten \cite{spattn2021} and BESAPU \cite{qinghua2023} for the following reasons.
ELSA prefetches the key matrix and speculates candidates for every query token through orthogonal projection before attention computation. 
Its preprocessing needs to access the query and key matrix in SRAM numerous times and consumes large power, while the SRAM access of the pruning processing in DEFA takes less than 0.1\% of the overall SRAM access and the pruning process in DEFA consumes extremely less power.
% Moreover, the reduced computation in DEFA is less than our pruning methods.26.6\%
SpAtten structurally removes the attention tokens and heads with low cumulative scores.
Compared to the coarse-grained pruning methods in SpAtten, DEFA performs fine-grained FWP and PAP on fmap pixels and sampling points.
Therefore, DEFA attains a higher pruning ratio with acceptable AP loss, which saves more computation and energy than SpAtten.
BESAPU bidirectionally speculates and approximately computes weakly related tokens with an out-of-order scheduler to save energy consumption.
However, 
the improvement of the energy efficiency resulting from the approximate computation highly depends on the ratio of weakly related tokens, which denotes the sparsity of attention, because the complex control logic consumes large energy.
% However, the improvement of the energy efficiency benefiting from the approximate computation highly depends on the sparsity of attention due to the complex control logic for processing weakly related tokens.
% only when BESAPU is high sparsity, the energy saving benefiting from the approximate computation of weakly related tokens can exceed the energy consumption of the complex control logic for BESAPU, which 
In contrast, fine-grained operator fusion and fmap reuse can save a large amount of memory access to enhance energy efficiency in MSGS without the restriction of sparsity.
Besides, the speculation in BESAPU only reduces the negative operations of weakly related tokens, while FWP and PAP in DEFA can remove all the computation of the redundant fmap pixels and sampling points, thus saving more energy.
Hence, DEFA achieves higher energy efficiency than BESAPU.

% The operator fusion and bounded range reuse save a large amount of SRAM access in MSGS and the FWP and PAP remove redundant fmap pixels and sampling points of MSGS input. 
% Therefore, the MSGS processing in DEFA are more efficient than the sparse matrix multiplication, the main procedure in the sparse attention works.
% Besides, the sparsity-detecting methods in the attention accelerators have more complex hardware implementations than the mask generation units in DEFA, which causes higher energy consumption.
% In addition, all the sparse matrix multiplications after FWP and PAP in DEFA are structural while 
% The proposed fine-grained operator fusion and fmap reuse save
% a large amount of SRAM access in MSGS without dependence on
% sparsity and consequently DEFA achieves higher energy efficiency.
\begin{table}[t]
    \small
    \centering
    \caption{Comparison with Other ASIC Platforms}
    \label{tab:V_Performance_Comparison}
    \renewcommand\arraystretch{1.3}
    % \resizebox{\textwidth}{!}{
	\begin{threeparttable}
        \begin{tabular}{l|c|c|c|c}
            \toprule
            % & JSSC'22\cite{PNNASIC_JSSC21} & VLSI'21\cite{PNNPU_VLSI21} & TCAD'23\cite{QuickFPS_TCAD23} & \textbf{FLNA} \\
            % & \cite{PNNASIC_JSSC21} & \cite{PNNPU_VLSI21} & \cite{QuickFPS_TCAD23} & \textbf{FLNA} \\
            & \tabincell{c}{\cite{elsa}\\ISCA'21} & \tabincell{c}{\cite{spattn2021}\\HPCA'21} & \tabincell{c}{\cite{qinghua2023}\\JSSC'22} & \textbf{DEFA} \\
            \hline
            Function & \multicolumn{3}{c|}{Attention}& \tabincell{c}{Deform\\Attn} \\
            \hline
            Technology(nm) & 40 & 40 & 28 & 40 \\
            \hline
            Area(mm$^2$) & 1.26 & 1.55 & 6.82 & 2.63\\
            \hline
            Frequency(MHz) & 1000 & 1000 & 500 & 400 \\
            \hline
            Precision & INT9 & INT12 & INT12 & INT12 \\
            % \hline
            % Throughput(GOPS)  & 737.3(8b) & 13400(fp32) & 760.8(8b) \\
            % \tabincell{c}{Peak Throughput\\(GOPS)}  & 737.3 & 13400 &  \\
            \hline
            Power(mW) & 969.4 & 294.0 & 272.8 & 99.8 \\
             \hline
            \tabincell{c}{Throughput(GOPS)} & 1088 & 360 & 522 & 418 \\
            \hline
            \tabincell{c}{Energy Effi.(GOPS/W)} & 1120 & 1224 & 1910 & 4187 \\
            % \hline
            % % \tabincell{c}{Energy Efficiency\\(TOPS/W)}  & 2.701 & 0.054 & 3.261 \\
            % \hline
            % Latency(ms) & 4.45$^{1}$ & 11.79$^{2}$ & \tabincell{c}{0.1$^{3}$\\ 3$^{4}$ \\ 40$^{5}$} & \tabincell{c}{\textbf{0.13-1.89}$^{4}$ \\ \textbf{0.88-1.59}$^{6}$ \\ \textbf{2.43}$^{7}$} \\
            \bottomrule
        \end{tabular}
        % \begin{tablenotes}
        %     \footnotesize
        %     \item[1] 320$\times$240 RGB-D images @ MSRA15 dataset\cite{MSRA15_CVPR15}.
        % \end{tablenotes}
	\end{threeparttable}
\end{table}

%% file: 6_conclusion.tex
\section{Conclusion}
We propose DEFA, the first algorithm-architecture co-design for efficient MSDeformAttn acceleration. 
On the algorithm level, we present FWP and PAP to effectively prune the fmap pixels and sampling points in MSGS so that the memory access and computation are significantly reduced. 
On the architecture level, DEFA adopts inter-level parallel processing for throughput enhancement. 
DEFA further utilizes fine-grained operator fusion and fmap reuse to alleviate memory footprint.
DEFA achieves speedup and energy saving over Nvidia RTX 2080Ti and 3090Ti, and higher energy efficiency than the SOTA attention accelerators. 